\documentclass[12pt]{article}
\usepackage{amsmath,amssymb}
\usepackage{epsfig}
\usepackage{graphicx}
\usepackage{a4}
\usepackage{latexsym}

\usepackage{color}
\usepackage{colordvi}
\usepackage{bm}

\graphicspath{{figure/}}
\DeclareGraphicsExtensions{.eps,.ps}

\usepackage{pslatex}
\usepackage{txfonts}
\usepackage[latin1]{inputenc}
\usepackage[T1]{fontenc}

\begin{document}
\setlength{\parskip}{0.2cm}
\setlength{\baselineskip}{0.535cm}

\begin{titlepage}
\noindent
DESY 08-185\\
SFB/CPP-08-99 \\
December 2008 
\begin{center}
\Large {\bf Determination of Strange Sea Distributions \\
from $\bm{\nu N}$ Deep Inelastic Scattering}

\vspace{1.2cm}
\large
S. Alekhin$^{\, a,b,}$\footnote{{\bf e-mail}: sergey.alekhin@ihep.ru},
S. Kulagin$^{\, c,}$\footnote{{\bf e-mail}: kulagin@ms2.inr.ac.ru},
R. Petti$^{\, d,}$\footnote{{\bf e-mail}: Roberto.Petti@cern.ch}
\vspace{1.0cm}

\normalsize
{\it $^a$Deutsches Elektronensynchrotron DESY \\
\vspace{0.1cm}
Platanenallee 6, D--15738 Zeuthen, Germany}\\
\vspace{0.5cm}
{\it $^b$Institute for High Energy Physics \\
\vspace{0.1cm}
142281 Protvino, Moscow region, Russia}\\
\vspace{0.5cm}
{\it $^c$Institute for Nuclear Research of the Academy of
Sciences of Russia \\
\vspace{0.1cm}
117312 Moscow, Russia}\\
\vspace{0.5cm}
{\it $^d$Department of Physics and Astronomy, University
of South Carolina \\
\vspace{0.1cm}
Columbia SC 29208, USA}
\vspace{0.5cm}
\vspace{0.5cm}

\large {\bf Abstract}
\vspace{-0.2cm}
\end{center}
We present an analysis of the nucleon strange sea extracted from a global 
Parton Distribution Function fit including the neutrino and anti-neutrino  
dimuon data by the CCFR and NuTeV collaborations, the inclusive charged lepton-nucleon 
Deep Inelastic Scattering and Drell-Yan data. The (anti-)neutrino induced dimuon analysis 
is constrained by the semi-leptonic charmed-hadron branching ratio
$B_\mu=(8.8\pm0.5)\%$, determined from the inclusive charmed hadron measurements performed 
by the FNAL-E531 and CHORUS neutrino emulsion experiments.
Our analysis yields a strange sea suppression factor 
$\kappa(Q^2=20~{\rm GeV}^2)=0.62\pm 0.04$,
the most precise value available, an $x$-distribution of total strange sea that is
slightly softer than the non-strange sea, and an asymmetry between strange and anti-strange
quark distributions consistent with zero (integrated over $x$ it is equal to
$0.0013 \pm 0.0009$ at $Q^2=20~{\rm GeV}^2$).
%
\vspace{1.5cm}
\end{titlepage}

\section{Introduction}

The strange quark ($s$) distribution in the nucleon is an important 
input for the QCD phenomenology since the contribution of the $s$-quarks to the hard cross sections 
is of the same order of magnitude as the non-strange quarks. 
The strange quark contribution is particularly important at small values of the 
parton momentum fractions $x$, where the quark distributions
are dominated by the sea.
In high-energy hadron collisions the region of $x\lesssim 0.1$
is crucial for the study 
of many processes and therefore an accurate determination of the strange 
sea is required for the interpretation of experimental data. 
For instance a small positive $s-\overline{s}$ asymmetry in the strange sea  
may help explain the anomaly in the weak mixing angle reported by the NuTeV 
experiment~\cite{Davidson:2001ji}. 
Inclusive cross sections are not very 
sensitive to the strange sea, since in this case the 
complementary contributions from strange 
and non-strange distributions are strongly 
anti-correlated. The strange sea is best
 constrained by the neutrino-nucleon deep-inelastic 
scattering (DIS) dimuon data. 
This process stems from the charged-current (CC) production of 
a charm quark, which semileptonically decays into a final state secondary muon. 
The charm quark production cross section involves
terms proportional to both the strange and the non-strange quark distributions. 
However, the contributions from $u$- and $d$-quarks are suppressed by the small quark-mixing
Cabibbo-Kobayashi-Maskawa (CKM) matrix elements. 
The most precise (anti-)neutrino dimuon data currently
available are by the CCFR and NuTeV 
collaborations~\cite{Bazarko:1994tt,Goncharov:2001qe,Mason:2007zz}.
In this paper we describe a determination of the strange sea distributions from 
a global parton distribution function (PDF) fit to the hard scattering processes, 
such as the inclusive charged-leptons DIS and Drell-Yan data, 
with the inclusion of the important CCFR and NuTeV dimuon data.
The analysis is performed in the next-to-next-to-leading-order    
(NNLO) QCD approximation for the PDF evolution 
and for the massless coefficient functions.
The next-to-leading-order (NLO) QCD corrections to the CC
heavy-quarks production cross section are taken into 
account. These corrections reduce theoretical uncertainties on the strange sea due to variations 
in the renormalization and factorization scale. 

The paper is organized as follows.
Section~2 presents the theoretical framework for (anti-)neutrino induced 
charm dimuon production. In Section~3 we discuss the result of our global fit and the  
dominant theoretical uncertainties in the extraction of (anti-)strange quark distributions.  
We also discuss the impact of the semileptonic charm quark branching ratio $B_\mu$ on the 
strange distributions and we present an updated value of this parameter. 
Comparisons of these results with the earlier determinations 
of the strange sea from the leading-order (LO) analysis of Ref.~\cite{Goncharov:2001qe},
the NLO analysis of Refs.~\cite{Bazarko:1994tt,Lai:2007dq}, and 
the NNLO analysis of Ref.~\cite{MSTW} are presented. Section~4 outlines future improvements 
in the determination of the strange sea distributions.

\section{Theoretical Framework}
\label{sec:theory} 

\begin{figure}[htb]
\centerline{\epsfxsize=5in\epsfysize=4.5in\epsfbox{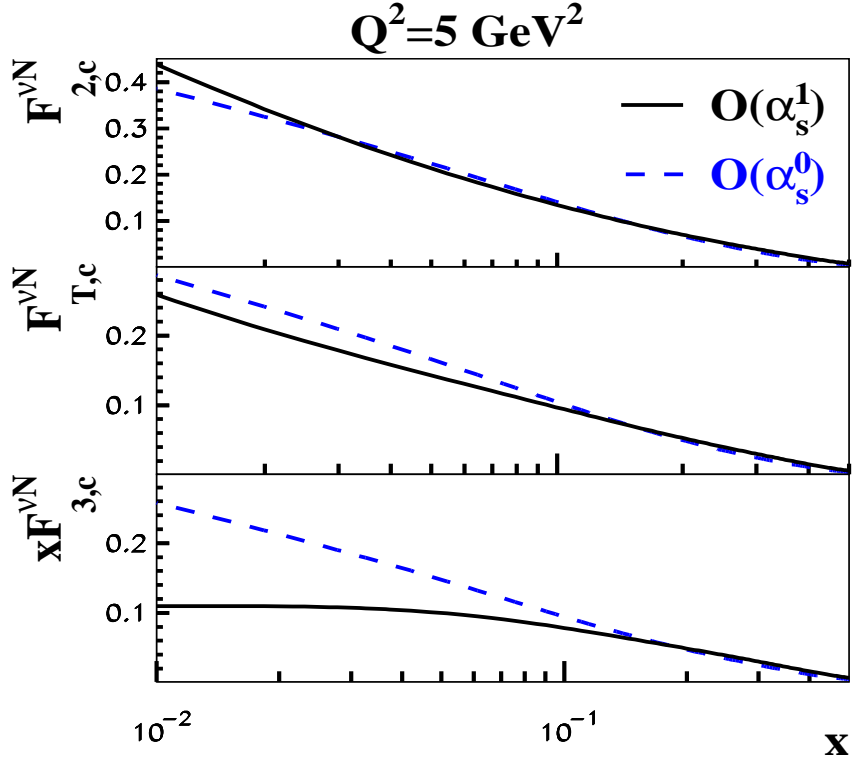}}
\caption{Comparison between LO (dashed) and NLO (solid) QCD approximations for 
charm quark production structure functions. The calculation is performed for  
neutrino interactions on isoscalar nucleons. 
\label{fig:sfc}}
\end{figure}

The differential cross section for charm quark production in CC (anti-)neutrino DIS 
off nucleon or nuclear target can be written as:  
\begin{eqnarray}
\frac{d\sigma
\genfrac{}{}{0cm}{2}{{\genfrac{}{}{0cm}{3}{(-)}{^\nu}}}{\rm charm}}
{dxdy}
=\frac{G_F^2ME}
{\pi(1+Q^2/M_W^2)^2}\left[\left(1-y-\frac{Mxy}{2E}\right)
F\genfrac{}{}{0cm}{2}{{\genfrac{}{}{0cm}{3}{(-)}{^\nu} }}{\rm 2,c~~~}(x,Q^2)
+\right. \nonumber\\
\left. +\frac{y^2}{2}
F\genfrac{}{}{0cm}{2}{{\genfrac{}{}{0cm}{3}{(-)}{^\nu} }}{\rm T,c~~~}(x,Q^2)
\genfrac{}{}{0cm}{1}{+}{(-)} 
y\left(1-\frac{y}{2}\right)
xF\genfrac{}{}{0cm}{2}{{\genfrac{}{}{0cm}{3}{(-)}{^\nu} }}
{\rm 3,c~~~}(x,Q^2)
\right],
\label{eqn:cs}
\end{eqnarray}
where $x$, $y$, and $Q^2$ are common DIS variables, $E$ 
is the (anti-)neutrino energy, $G_F$ is the Fermi constant,  
$M$ and $M_W$ are the nucleon and $W$-boson masses, respectively,  
and $F_{2,T,3}$ are the corresponding structure functions (SFs).
The nuclear data are usually presented in terms of an isoscalar target nucleon,
which is the average over proton and neutron targets. 
For an isoscalar nucleon, 
assuming the usual isospin relations between the proton and
neutron quark distributions, we have in the LO QCD approximation: 
\begin{eqnarray}
F_{\rm 2,c}^{^{\genfrac{}{}{0cm}{2}{(-)}{^\nu}} {^N}}(x,Q^2)
=2\xi \left[\left \vert V_{cs} \right \vert ^2
\genfrac{}{}{0cm}{1}{(-)}{^{^{\displaystyle s}}}(\xi,\mu^2) 
+ \left \vert V_{cd} \right \vert ^2
\frac{\genfrac{}{}{0cm}{1}{(-)}{^{^{\displaystyle u}}}(\xi,\mu^2)
+\genfrac{}{}{0cm}{1}{(-)}{^{^{\displaystyle d}}}(\xi,\mu^2)}{2} \right],
\nonumber \\
F_{\rm T,c}^{^{\genfrac{}{}{0cm}{2}{(-)}{^\nu}} {^N}}
=\genfrac{}{}{0cm}{1}{+}{(-)} xF_{\rm 3,c}^{^{\genfrac{}{}{0cm}{2}{(-)}{^\nu}}
{^N}}
=\frac{x}{\xi}F_{\rm 2,c}^{^{\genfrac{}{}{0cm}{2}{(-)}{^\nu}} {^N}},
~~~~~~~~~~~~~~~~~~~~~~~~
\label{eqn:sf}
\end{eqnarray}
where $u,d,s$ are the light quark distributions in the proton,
$\xi=x(1+m_{\rm c}^2/Q^2)$ is the slow-rescaling variable appearing  
in the kinematics of $2\to2$ parton scattering 
with one massive particle in the final state~\cite{Barnett:1976ak},  
and $m_{\rm c}$ is the charm quark mass. The values of the CKM 
matrix elements
$V_{cs}= 0.97334$ and $V_{cd}=0.2256$~\cite{PDG08} suggest that
the strange quark contribution dominates 
the cross section of Eq.~(\ref{eqn:cs}) at small $x$. 
The factorization scale $\mu$ is usually set to either $Q$ or 
$\sqrt{Q^2+m_{\rm c}^2}$. The sensitivity to a particular choice of
$\mu$ gives an idea about the impact of higher-order QCD corrections.
In the NLO QCD approximation 
the structure functions of Eq.~(\ref{eqn:sf}) get an additional $O(\alpha_{\rm s})$ 
contribution from the gluon-radiation and gluon-initiated 
processes~\cite{Gottschalk:1980rv}.
In Fig.~\ref{fig:sfc} we compare the structure functions for charm production
calculated in the NLO and LO approximations.
The magnitude of NLO corrections 
rises at small $x$, giving the largest effect in the case of $xF_3$. 
For realistic kinematics, the NLO corrections to 
Eq.~(\ref{eqn:sf}) substantially cancel out
in the difference between neutrino and anti-neutrino cross sections.
In practice higher-order QCD corrections affect mainly the 
$C$-even combination $s+\bar s$. 
We calculate the QCD-evolution of PDFs in the NNLO 
approximation~\cite{Moch:2004pa}. 
However, a fully consistent NNLO calculation of the structure functions 
in Eq.~(\ref{eqn:cs}) is currently not possible, since
the NNLO coefficient functions for charm quark 
production are not available.
The contribution to NNLO corrections from the soft-gluon re-summation 
has been calculated in Ref.~\cite{Corcella:2003ib} and is 
significant only at large values of $x$. Therefore, 
we do not include these corrections in our analysis.  
In general, the NNLO corrections are expected to be small compared to the 
uncertainties of experimental data, as one can infer from the typical 
magnitude of NLO corrections.

We do not consider power corrections to the SFs of Eq.~(\ref{eqn:sf}).
The target mass corrections of Ref.~\cite{Kretzer:2003iu} are marginal in the 
region of $x<0.3$ covered by the CCFR and NuTeV dimuon data.   
The dynamical high-twist contributions to the charm production SFs are unknown. 
We estimate their effect by applying a simple rescaling for the quark charge 
to the phenomenological twist-4 terms extracted from the inclusive $\nu N$ 
cross-sections~\cite{AKP07,Alekhin:2008ua}. Following 
this procedure we find that 
the impact of these corrections is negligible.

\begin{figure}[ht]
\centerline{\epsfxsize=6.5in\epsfysize=3.5in\epsfbox{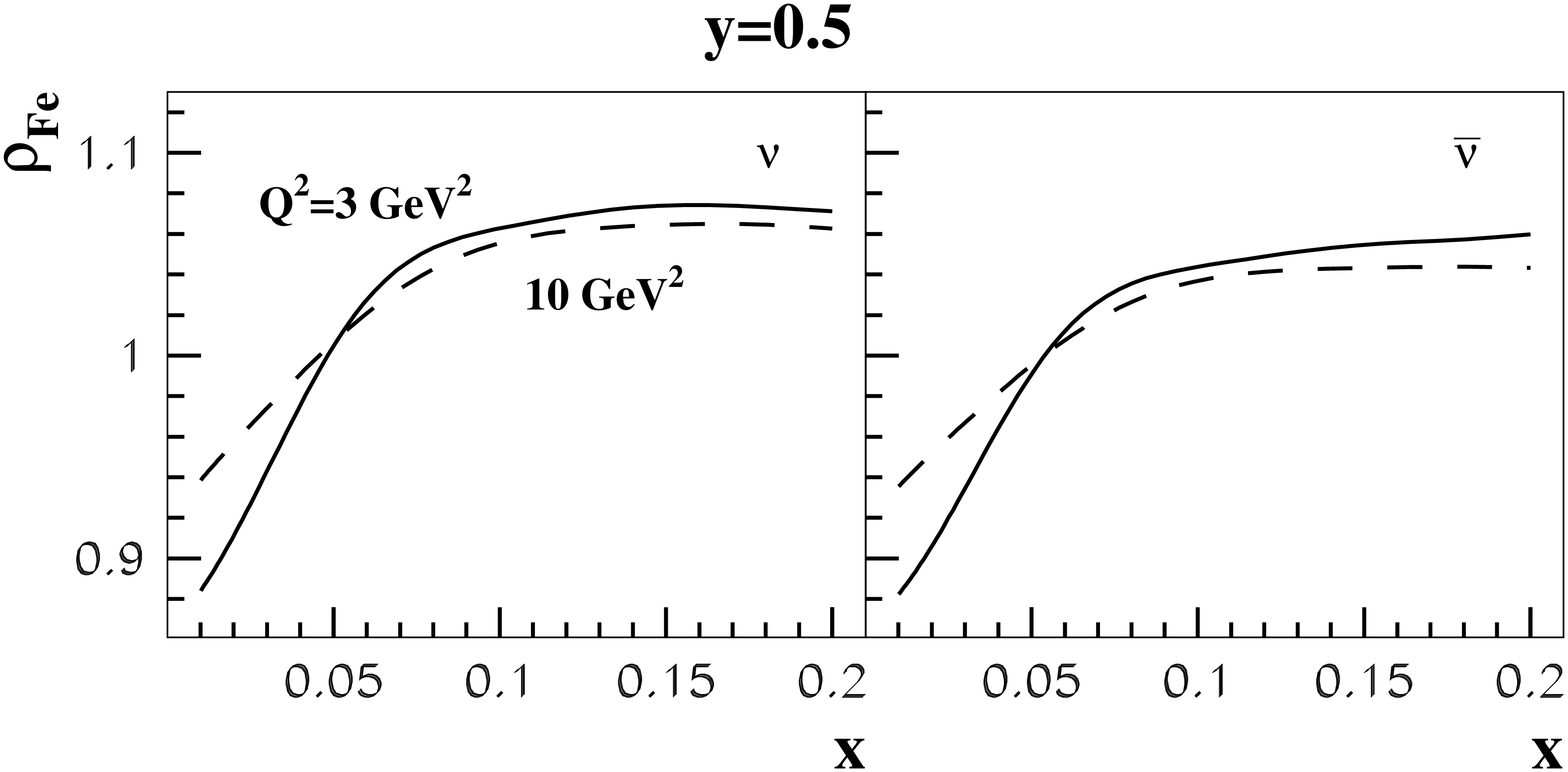}}
\caption{Ratio of differential cross sections for iron and 
isoscalar nucleon, $\rho_{\rm Fe}$, 
in neutrino (left panel) and anti-neutrino (right panel) interactions~\cite{KP04,KP07}. 
The solid (dashed) curve corresponds to $Q^2=3\,(10)\,{\rm GeV}^2$.
The inelasticity, $y$, is fixed at 0.5.}
\label{fig:nc}
\end{figure}

Data from the CCFR and NuTeV experiments were collected on iron target. 
We apply nuclear corrections to Eq.~(\ref{eqn:sf}) using the calculation of 
Ref.~\cite{KP04,KP07}. This calculation takes into account a number of different effects
including the Fermi motion and binding, neutron excess, 
nuclear shadowing, nuclear pion excess and 
the off-shell correction to bound nucleon SFs. The model of Ref.~\cite{KP04} provides 
a good description of the nuclear EMC effect as measured in charged-lepton DIS over a
wide range of nuclear targets, from deuterium to lead. 
In Ref.~\cite{KP07} this approach was extended to describe the (anti-)neutrino interactions
with nuclei. The model predicts that nuclear corrections to the
neutrino-nucleon structure functions
are different from those for charged-lepton interactions.
Furthermore, nuclear effects for the case of (anti-)neutrino scattering
depend on the SFs
type ($F_2$ vs. $xF_3$) and on the specific $C$-parity and isospin states.  
Fig.~\ref{fig:nc} shows the nuclear corrections for neutrino and anti-neutrino 
differential cross sections calculated for an iron target.

Electroweak corrections including the one-loop terms
are calculated in Ref.~\cite{Arbuzov-Bardin}, 
within the framework of the parton model, in a factorized form. 
In this approach 
the initial quark mass singularities of the QED diagrams are subtracted
within the $\overline{\rm MS}$ scheme and included into the PDFs, which 
absorb all electroweak corrections. It is interesting to note that 
the electroweak and nuclear corrections are similar in magnitude
in certain kinematic regions. Since the dimuon data released by the
NuTeV and CCFR collaborations have already been corrected for electroweak effects 
according to an earlier calculation of Ref.~\cite{Bardin-Dokuchaeva}, we do not 
apply such corrections in our fit.  
  
\begin{figure}[ht]
\centerline{\epsfxsize=6in\epsfysize=3.5in\epsfbox{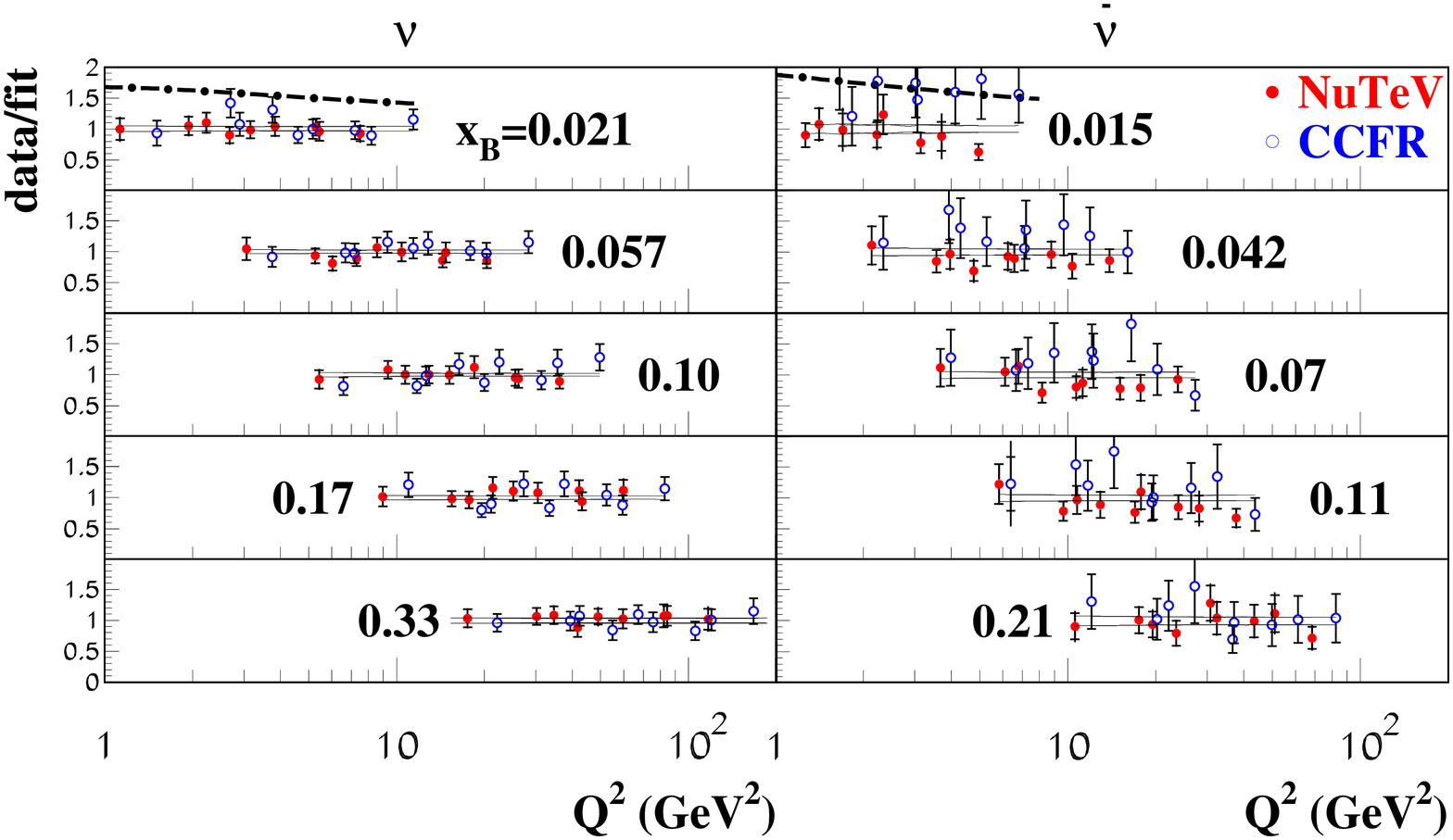}}
\caption{Pulls of CCFR and NuTeV dimuon data with respect to our fit (left panel: neutrino, 
right panel: anti-neutrino).
The solid lines represent a $\pm 1\sigma$ band for the fitted model. The dashed dots 
illustrate the impact of an (anti-)strange sea enhancement
on the (anti-)neutrino cross sections at small $x$.
}
\label{fig:pull}
\end{figure}

In the LO the dimuon cross section 
is related to the corresponding cross section for charmed-quark production as:  
\begin{equation}
\frac{d\sigma_{\mu\mu}}{dxdydz} =   
\frac{d\sigma_{\rm charm}}{dxdy} 
\sum_{\rm h} f_{\rm h} D_{\rm c}^h(z) Br(h\to \mu X),   
\label{eqn:frag}
\end{equation}
where $f_{\rm h}$ is the fraction of the charmed hadron $h$,   
$D_{\rm c}^h(z)$ is the fragmentation function of the charm quark 
into a given charmed hadron $h=D^0,D^+,D_s^+,\Lambda_{\rm c}^+$
carrying a fraction $z$ of the charm quark momentum, 
and $Br(h\to \mu X)$ is the corresponding inclusive branching ratio 
for the muon decays (note: the normalization $\sum f_{\rm h} =1$). 
In the NLO the coefficient functions entering the SFs 
calculation depend, in general, on $z$ as well.    
The charm fragmentation function $D_{\rm c}(z)$ defines the energy of the 
outgoing charmed hadron and, in turn, of the secondary muon    
produced in the semileptonic decays. 
Typically, a minimal energy $E_\mu^0$ is required for the muons identified 
experimentally, in order to suppress the background from light-meson semileptonic decays. 
Assuming a universal $D_{\rm c}(z)$ for all charmed hadrons and integrating over $z$,    
Eq.~(\ref{eqn:frag}) reads:  
\begin{equation}
\frac{d\sigma_{\mu\mu}(E_\mu > E_\mu^0)}{dxdy}=
\eta_\mu B_\mu
\frac{d\sigma_{\rm charm}}{dxdy},
\label{eqn:cut}
\end{equation}
where $\eta_\mu$ is the acceptance correction accounting for the cut 
$E_\mu > E_\mu^0$, and $B_\mu=\sum f_{\rm h} Br(h\to\mu X)$ is the effective 
semileptonic branching ratio. We use the values of $\eta_\mu$ evaluated by 
the NuTeV and CCFR collaborations~\cite{Mason:2006qa}, which are based 
on the NLO calculations of Ref.~\cite{disco} and on the Collins-Spiller~\cite{ColSpi} 
fragmentation function. The parameter $\varepsilon_{\rm c}$, which  
defines the shape of $D(z)$ in the Collins-Spiller model,  
is fixed at 0.6. This value corresponds to the best fit value obtained in the 
NuTeV analysis of Ref.~\cite{Mason:2006qa}.     

The charmed fractions $f_{\rm h}$ depend on the incoming neutrino energy.
This fact can be explained by the contributions from quasi-elastic $\Lambda_c$ 
and diffractive $D_s^\pm$ production.
Furthermore, the values of $f_{\rm h}$ are different for
neutrino and anti-neutrino beams since in the second case 
no quasi-elastic $\bar{\Lambda}_{\rm c}$ production is present, but    
the relative rate of diffractive $D_s$ production is about a factor of 
two larger.   
These two contributions are significant mainly at low energies and 
they would not affect the value of $B_\mu$ at $E_\nu>40$ GeV. 
Measurements of $f_{\rm h}$ and $B_\mu$ in neutrino interactions were 
performed by the E531~\cite{Ushida:1988rt,Bolton:1997pq}
and CHORUS~\cite{KayisTopaksu:2005je,DiCapua08} experiments using the emulsion
detection technique.
A value of $B_\mu=9.19\pm0.94\%$
was obtained in Ref.~\cite{Bolton:1997pq} by combining
the E531 data on $f_{\rm h}$ in the energy range $E>30$\ GeV, which is
relevant for the analysis of the NuTeV and CCFR dimuon data,
with the charmed-hadron semileptonic branching ratios.
The dominant source of uncertainty in this determination of $B_\mu$ is
related to the uncertainties on the charmed fractions $f_{\rm h}$.


A complementary determination of $B_\mu$ can be obtained from an analysis 
of dimuon data by performing a simultaneous fit of $B_\mu$ with other 
parameters~\cite{Bazarko:1994tt,Goncharov:2001qe}.  
In such an approach the absolute value of the dimuon cross section cannot 
directly constrain the (anti-)strange quark sea. 
This contribution is rather defined by the $Q^2$-slope of the cross section, 
which is sensitive to the parton distributions through the 
QCD evolution equations.
For anti-neutrinos the slope is driven mainly by the anti-strange sea,  
with a small contribution from gluons coming from the NLO corrections. 
In the neutrino case the non-strange quarks contribute as well.
Once the (anti-)strange distributions are constrained by the $Q^2$-slopes,
the parameter $B_\mu$ is determined by the absolute value of the dimuon 
(anti-)neutrino cross section. The value of $B_\mu$ obtained 
from this global fit can then be compared with the direct measurements from 
the emulsion experiments in order to check the self-consistency between the 
$Q^2$-slope and the absolute normalization of dimuon data.   

\section{Results}
\label{sec:results} 

\subsection{Constraints from CCFR and NuTeV Dimuon Data}
\label{sec:freeBmu}

We determine the strange sea distributions from a global PDF fit to the CCFR and NuTeV 
dimuon data of Refs.~\cite{Bazarko:1994tt,Mason:2007zz},  
combined with the inclusive charged-leptons DIS and the Drell-Yan
cross sections used in the earlier fit of Ref.~\cite{Alekhin:2006zm}.
The $x$-dependence of the strange and anti-strange quark distributions is  
parametrized independently using a model similar to that used for other PDFs: 
\begin{equation}
{\genfrac{}{}{0cm}{1}{~~(-)}{^{^{\displaystyle x~s}}}}(x,Q_0^2)
=\genfrac{}{}{0cm}{1}{(-)}{^{^{\displaystyle ~A_{s_{~}}}}}
x^{^{\genfrac{}{}{0cm}{2}{(-)}{~a_s}}}(1-x)
^{^{\genfrac{}{}{0cm}{2}{(-)}{~b_s}}}
\label{eqn:form}
\end{equation}
at the starting value of the QCD evolution $Q_0^2=9~{\rm GeV}^2$. 
This functional form is flexible enough to describe the data.  
We do not observe any significant improvement in the quality of our fit 
by adding a polynomial factor to Eq.~(\ref{eqn:form}). 
The low-$x$ exponents $a_s$ and $\overline{a}_s$
are assumed to be the same as the one for the non-strange sea,  
since the existing dimuon data are not sensitive to them.   
The remaining parameters in Eq.~(\ref{eqn:form}) are extracted  
simultaneously with the non-strange PDF parameters, which   
essentially coincide with the ones obtained in Ref.~\cite{Alekhin:2006zm}.

\begin{table}\begin{center}
\begin{tabular}{c|c|c} \hline
Parameter & Free $B_\mu$ & Constrained $B_\mu$ \\ \hline\hline 
$A_s$ & $0.086\pm 0.007$ & $0.088\pm 0.005$ \\
$a_s$ & $-0.220\pm 0.004$ &  $-0.220\pm 0.004$\\
$b_s$ & $7.7\pm 1.0$ & $7.5\pm 0.5$\\
${\overline A}_s$ & $0.083\pm 0.008$ & $0.085\pm 0.006 $\\
${\overline a}_s$ & $-0.220\pm 0.004$ &  $-0.220\pm 0.004$ \\
${\overline b}_s$ & $8.0\pm 0.4$ & $7.9\pm 0.4$ \\ \hline   
$m_{\rm c}$ (GeV) & $1.31\pm0.11$ & $1.32\pm0.11$   \\ 
$B_\mu$ (\%) & $9.1\pm1.0$ & $8.80\pm 0.45$   \\ \hline 
\end{tabular}
\caption{Our results for the strange sea and charm production parameters. 
Central column: variant of the fit in which $B_\mu$ is extracted from the 
CCFR and NuTeV dimuon data only; 
right column: variant of the fit with $B_\mu$ constrained by emulsion experiments.} 
\label{tab:pars}
\end{center}
\end{table}

\begin{figure}[ht]
\centerline{\epsfxsize=6.5in\epsfysize=3.5in\epsfbox{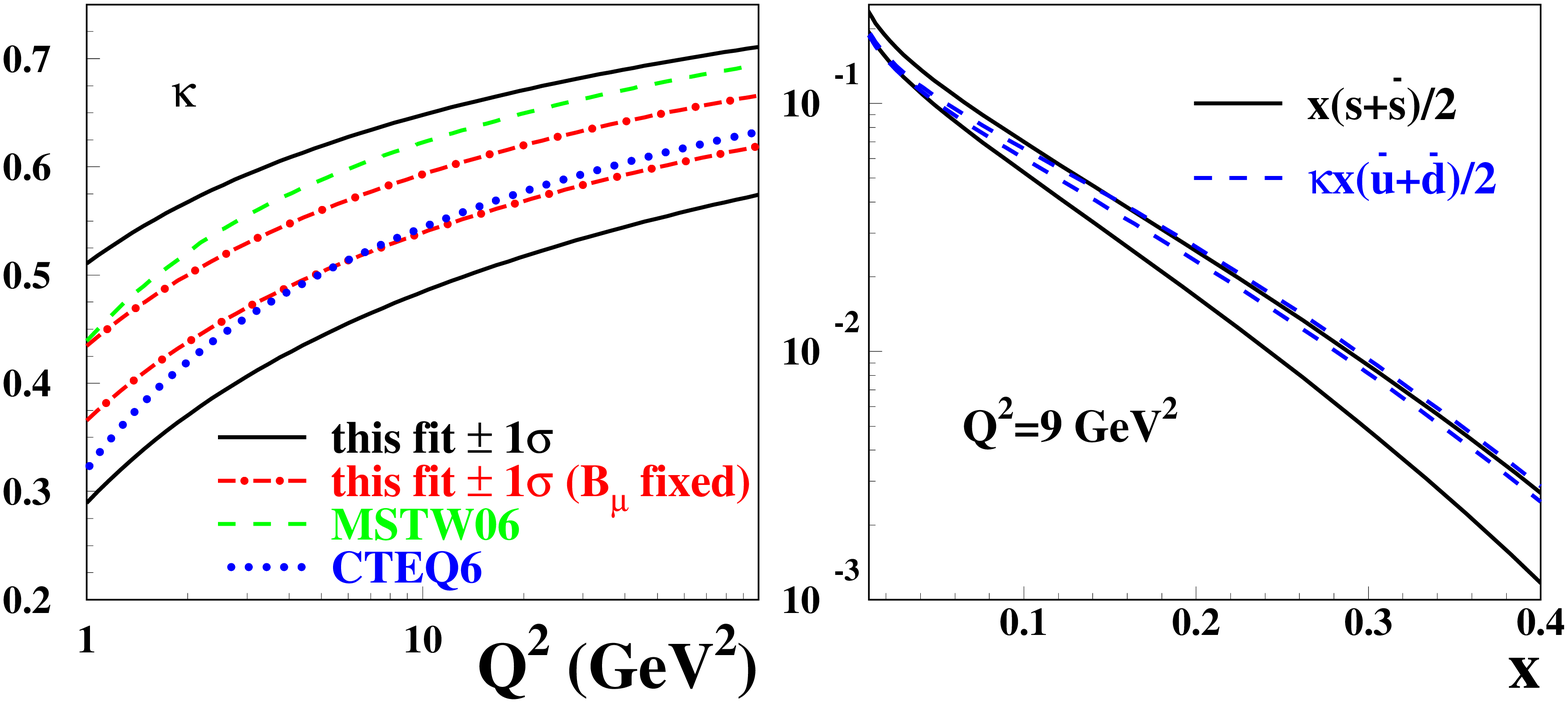}}
\caption{Left panel:
The $\pm 1\sigma$ band for the strange sea suppression factor
$\kappa$ obtained in our fit (solid lines) compared to the determinations 
by the MSTW (dashes) and CTEQ (dashed dots) collaborations. 
The dotted lines represent the corresponding band after fixing the 
value of $B_\mu$ to the central value obtained in our fit, 9.1\%.  
Right panel: The $\pm 1\sigma$ band for the $C$-even combination of the 
strange sea distributions determined in our fit (solid lines)
compared to the non-strange one scaled by $\kappa$ (dashes).
\label{fig:ssup}}
\end{figure}

Our results for the strange sea parameters are given in 
Table~\ref{tab:pars}. The quoted uncertainties  
include both statistical and systematic uncertainties in the data and 
take into account correlations in the latter where available.  
We obtain values of $\chi^2$ of 63 and 38 for the CCFR and NuTeV data 
sets, which both have 89 data points. It must be noted that, 
due to statistical correlations between data points, the effective 
number of degrees of freedom for the NuTeV data is about 40, which is 
consistent with our $\chi^2$ value. 

The ratio of CCFR and NuTeV data with respect to the fit model is  
given in Fig.~\ref{fig:pull}. Data from both experiments
are consistent and are in agreement with our fit in the whole 
kinematic range. Although the CCFR anti-neutrino data is higher than the model 
at small $x$, this discrepancy is within the uncertainties. 
The dashed curves in Fig.~\ref{fig:pull} illustrate the effect of 
increasing the strange sea normalization parameters $A_s$ and $\overline{A}_s$
by 0.03~\footnote{We choose a shift corresponding to several standard deviations 
for illustration purpose.}. One can see that both the normalization and 
the $Q^2$-slope of the fitted model change with the strange sea normalization. 
If we model the energy dependence of $B_\mu$ by a linear 
function, the corresponding slope obtained from the fit is comparable to zero 
within uncertainties. 
We also do not observe any significant difference between the 
values of $B_\mu$ obtained independently from the neutrino and anti-neutrino data sets:    
$9.4\pm1.1\%$ and $8.9\pm2.2\%$, respectively.  
The neutrino-antineutrino and energy-averaged value of  
$B_\mu=9.1\pm1.0~\%$ obtained in our fit is in good agreement with the results
of Ref.~\cite{Bolton:1997pq}.
%
We also extract the charm quark mass $m_{\rm c}$ 
from the data. We obtain a value $m_{\rm c}=1.31\pm0.11~{\rm GeV}$, 
which is in agreement with the world average $\overline{MS}$ value 
$m_{\rm c}=1.27^{+0.07}_{-0.11}~{\rm GeV}$~\cite{PDG08}.

\begin{figure}[ht]
\centerline{\epsfxsize=6.5in\epsfysize=3.5in\epsfbox{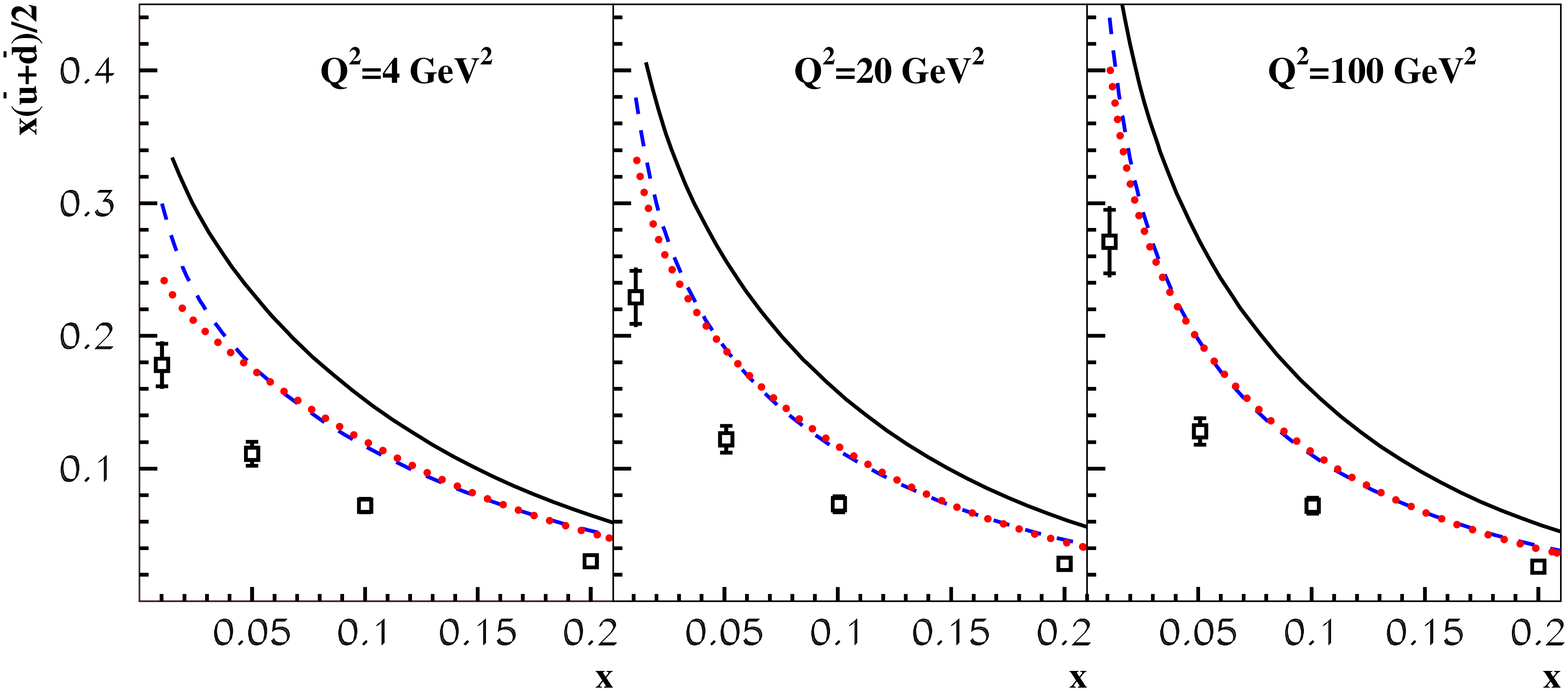}}
\caption{The non-strange sea distribution obtained in the NLO QCD fit 
of Ref.~\protect{\cite{Bazarko:1994tt}}
by the CCFR collaboration (solid line) in comparison
with the ones from MSTW06 (dashes) and CTEQ6 (dots) PDFs. 
The points give the corresponding strange-sea distribution extracted by 
CCFR~\protect{\cite{Bazarko:1994tt}}.
\label{fig:ccfr}}
\end{figure}

\begin{figure}[ht]
\centerline{\epsfxsize=6.5in\epsfysize=3.5in\epsfbox{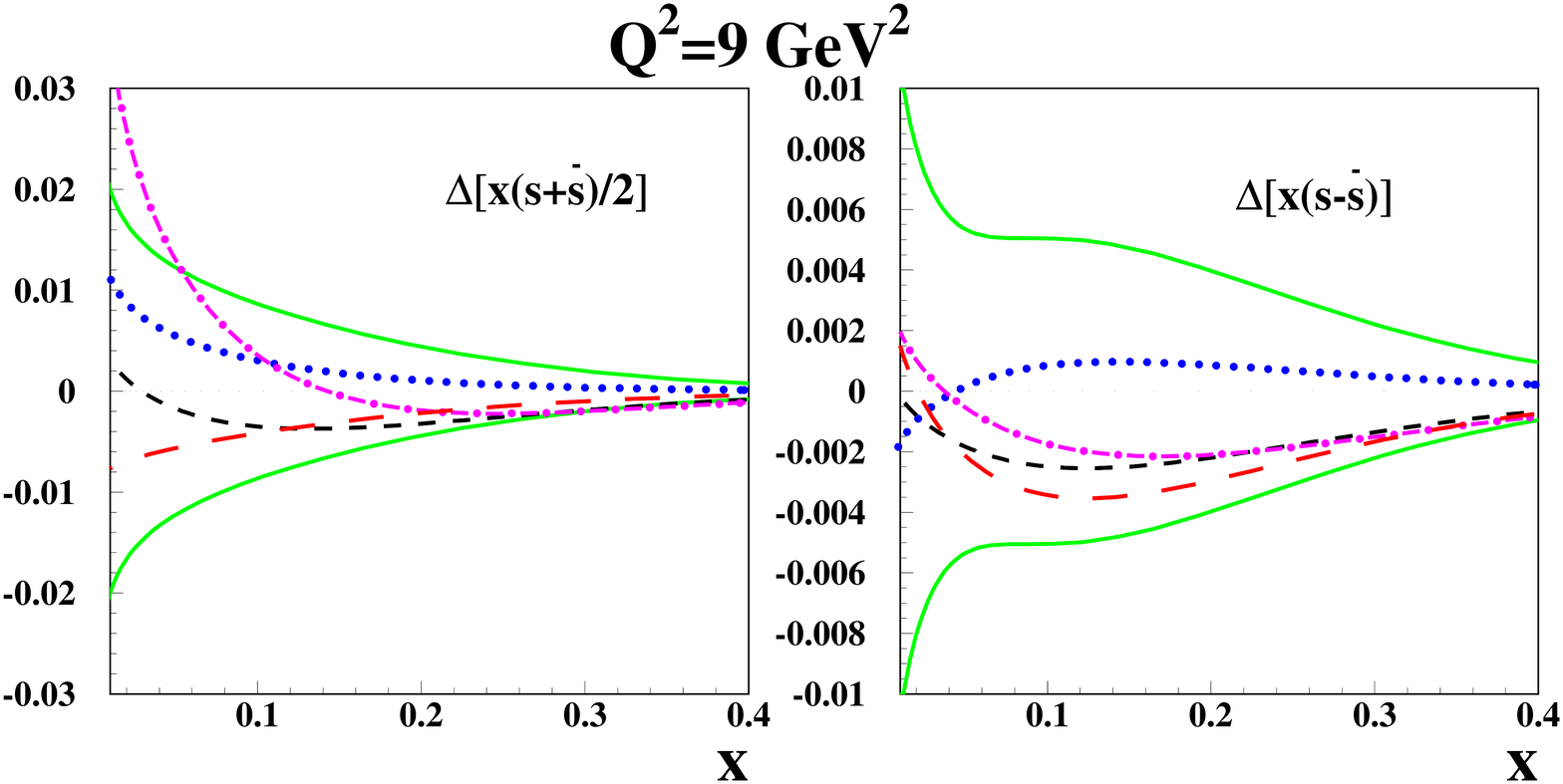}}
\caption{Sensitivity of the strange sea distribution to 
various corrections and settings of the fit. 
Left panel: The shifts in $C$-even $s+\bar s$ 
distribution due to the NLO QCD corrections 
to the charm quark production coefficient functions (dashed-dotted curve),
the variation of the QCD scale $\mu$ from $\sqrt{Q^2+m_{\rm c}^2}$ to 
$Q^2$ (dots), the NNLO corrections to the QCD evolution and 
the massless coefficient functions (long dashes), and  
the nuclear corrections (short dashes). The solid lines give 
the $\pm 1\sigma$ uncertainty band from our fit.  
Right panel: The same for the  $C$-odd distribution $s-\bar s$.}
\label{fig:corr}
\end{figure}

The strange sea suppression factor 
\begin{equation}
\kappa(Q^2)=\frac{\int_0^1 x\left[ s(x,Q^2)+\overline{s}(x,Q^2)\right]dx}
{\int_0^1 x\left[ \overline{u}(x,Q^2)+\overline{d}(x,Q^2)\right] dx},
\end{equation}
calculated with the PDFs obtained from our fit is given in Fig.~\ref{fig:ssup}.
The momenta carried by all sea quark flavors rise in the same way with $Q^2$, 
due to the QCD evolution. Therefore, the suppression factor $\kappa$ also increases 
with $Q^2$. We obtain $\kappa(20~{\rm GeV}^2)=0.59\pm 0.08$.
The uncertainty in the strange sea normalization parameters is correlated with the one 
on $B_\mu$. If we fix $B_\mu$ at the central value obtained in our fit,    
we observe a reduction by a factor of 3 in the uncertainty on $\kappa$.  
Our value of $\kappa$ is bigger 
than that obtained in the NLO QCD analysis of the
CCFR dimuon data~\cite{Bazarko:1994tt}, 
$\kappa(20~{\rm GeV}^2)=0.48\genfrac{}{}{0cm}{1}{+0.06}{-0.05}$. 
This difference occurs since the non-strange 
sea quark distributions used in Ref.~\cite{Bazarko:1994tt} 
are larger than those of both our fit and   
other modern sets of PDFs (Fig.~\ref{fig:ccfr}). However,
the strange sea from our fit 
is consistent with that of Ref.~\cite{Bazarko:1994tt}. 
The values of $\kappa$ calculated using the CTEQ6~\cite{Pumplin:2002vw} and 
MSTW06~\cite{Martin:2007bv} PDF sets agree with our determination within 
the uncertainties. 
The value of $\kappa$ preferred by the 
combined data on the vector meson electro-production in the analysis 
of Ref.~\cite{Goloskokov:2006hr} is also consistent with our 
determination.
The strange sea distribution obtained in our fit is somewhat softer 
than the non-strange one (see Fig.~\ref{fig:ssup}). 
Due to the NLO corrections to the charmed-quark production coefficient
functions the strange sea distributions are enhanced at
small $x$ (Fig.~\ref{fig:corr}).
If we do not take into account such corrections, we obtain a smaller
value of $\kappa(20~{\rm GeV}^2)=0.55\pm0.13$.
This effect is consistent with the difference between the values of $\kappa$
obtained in the NLO fit of Ref.~\cite{Bazarko:1994tt} and in the LO fit 
of Ref.~\cite{Goncharov:2001qe}. The variation of the strange sea due to 
a change of the QCD scale $\mu$ from $\sqrt{Q^2+m_{\rm c}^2}$ to $Q$ 
is smaller than the one due to the NLO correction to the charmed-quark 
production coefficient functions. This result indicates  
our fit is stable with respect to the higher-order QCD corrections.

\begin{figure}[ht]
\centerline{\epsfxsize=6in\epsfysize=3.5in\epsfbox{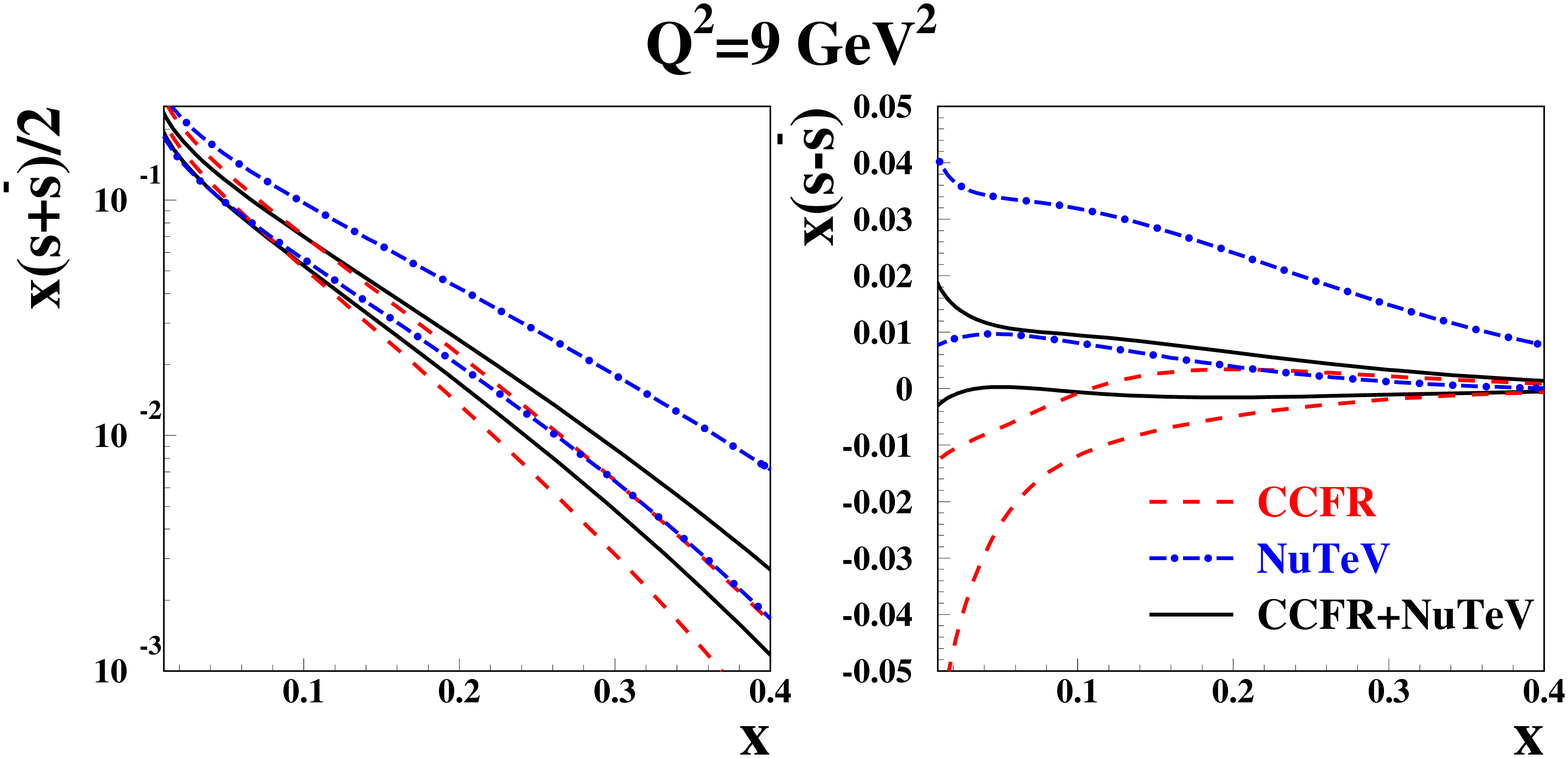}}
\caption{Left panel: $\pm 1\sigma$ bands for 
the C-even strange distribution $s+\bar s$ as obtained from   
the combined CCFR and NuTeV data (solid lines), from the CCFR data only (dashes), 
and from the NuTeV data only (dashed-dotted curves). Right panel: The same for the C-odd 
distribution $s-\bar s$.
\label{fig:ssea}}
\end{figure}

In a variant of the fit with only the NuTeV data the strange sea is somewhat 
enhanced with respect to the combined CCFR and NuTeV fit 
(see Fig.~\ref{fig:ssea}).
The value $B_\mu=7.2\pm1.7~\%$ obtained in this case is correspondingly
smaller than those from both the combined fit and the analysis of 
Ref.~\cite{Bolton:1997pq}.
Although the discrepancy is at the level of $1\sigma$, it might indicate a certain 
inconsistency between the $Q^2$-slope and the absolute normalization of the NuTeV data.
In a variant of the fit with only CCFR data  
we get $B_\mu=9.7\pm1.1~\%$, which is more consistent with 
the results from emulsion experiments. The strange sea determined from 
the CCFR data is somewhat smaller than the one from the combined fit.   
The strange sea charge asymmetry preferred by the NuTeV data is positive at all $x$ 
values and is consistent with the analysis by the NuTeV collaboration~\cite{Mason:2007zz}. 
However, the CCFR data prefer negative charge asymmetry, so that the combined CCFR and NuTeV 
value is consistent with zero at the initial scale $Q_0^2=9~{\rm GeV}^2$ (Fig.~\ref{fig:ssea}). 
Once we impose the constraint $s(x)=\overline{s}(x)$,  
we observe an increase of $\chi^2$ limited to about one unit.  
The variant of fit with the constraint 
$\int^1_0[s(x)-\overline{s}(x)]dx=0$ 
imposed also does not yield statistically significant increase in the value 
of $\chi^2$.


The strange sea asymmetry rises with $Q^2$~\cite{Catani:2004nc} because of the 
NNLO corrections. However, even taking into account such an effect, it remains 
consistent with zero within uncertainties in a wide range of $Q^2$. 
In particular, at the reference scale $Q^2=20~{\rm GeV}^2$ we obtain
$S^-=\int^1_0x[s(x)-\overline{s}(x)]dx=0.0010(13)$.
The value of $S^-$ is sensitive to $B_\mu$: if we fix $B_\mu$ the 
uncertainty on $S^-$ is reduced by about a factor of 2. 
The choice of the QCD scale $\mu$ and the details of the high-order QCD corrections 
for the non-strange quark contributions to the charm SFs also affect $S^-$ 
(Fig.~\ref{fig:corr}). Changing the QCD scale $\mu$ from $\sqrt{Q^2+m_{\rm c}^2}$ to $Q$ 
leads to a significant enhancement of the strange-anti-strange 
asymmetry at $x \sim 0.15$.
The NNLO corrections to the QCD evolution and to the massless coefficient 
functions change the $Q^2$-slope of the neutrino-nucleon DIS cross section.
As a result, the strange sea distributions extracted from the fit, which are 
sensitive to this slope, are modified and the value of the strange asymmetry decreases. 
The nuclear corrections, discussed in Section~2, further reduce the asymmetry at $x\sim 0.1$.   
Each of these factors change the value of the asymmetry within $0.5\sigma$.  
A combination of the effects discussed can, in principle, explain the difference 
between our result and those of Refs.\cite{Lai:2007dq,MSTW}, in which a  
positive $s-\bar s$ asymmetry at the level of 1-2$\sigma$ was reported.

\begin{figure}[ht]
\centerline{\epsfxsize=5.5in\epsfysize=3.5in\epsfbox{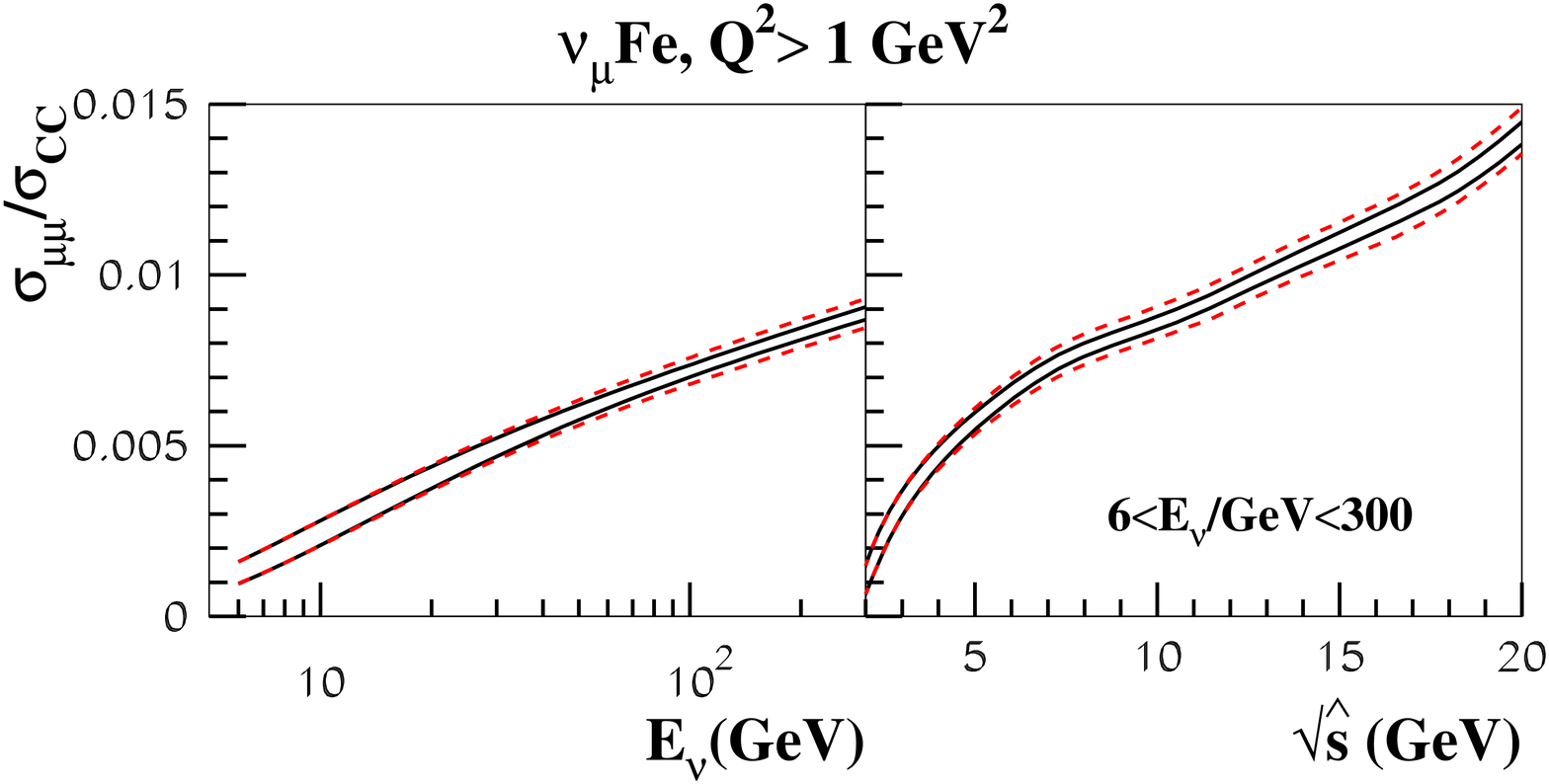}}
\caption{Left panel: The 
$\pm 1\sigma$ band for the ratio of the integral dimuon to the inclusive CC 
$\nu$Fe cross sections as a function of the neutrino energy calculated 
using the results of our fit (solid curves). The  
charm production cross section ratio rescaled by the value  
$B_\mu=8.8\%$ is also given for comparison (dashes). 
Right panel: The same 
for $d\sigma/d\hat{s}$, integrated over the neutrino energy spectrum 
of the NOMAD experiment~\cite{NOMAD} in the range of $6\div300~{\rm GeV}$.
A cut $Q^2>1~{\rm GeV}^2$ is imposed in both cases.}
\label{fig:rsig}
\end{figure}

\subsection{Impact of E531 and CHORUS Emulsion Data}
\label{sec:bmu} 

\begin{table}\begin{center}
\begin{tabular}{l|cc|cc} \hline
Measurement    & \multicolumn{2}{c|}{$E_\nu > 5$ GeV} & \multicolumn{2}{c}{$E_\nu > 30$ GeV} \\ \hline\hline   
CHORUS direct~\cite{KayisTopaksu:2005je} && $7.30\pm0.82$ && $8.50\pm1.08$  \\ 
CHORUS charmed fractions~\cite{DiCapua08} && $9.11\pm0.93$ &&   \\ 
E531 charmed fractions~\cite{Bolton:1997pq} && $7.86\pm0.49$ && $8.86\pm0.57$  \\ \hline\hline 
Weighted average && $7.94\pm0.38$ && $8.78\pm0.50$  \\ \hline
\end{tabular}
\caption{   
Semileptonic branching ratio $B_\mu(\%)$ from direct measurements in the E531 and CHORUS 
emulsion experiments. The last row corresponds to our weighted average. 
\label{tab:Bmu}
}
\end{center}
\end{table}

\begin{table}\begin{center}
\begin{tabular}{l|c|c|c|c} \hline
Energy (GeV) &  $5<E_\nu < 20$ & $20 < E_\nu<40$ & $40<E_\nu<80$ & $E_\nu>80$ \\ \hline\hline   
$B_\mu$ (\%) &  $6.33\pm1.05$ & $7.46\pm0.80$ & $8.68\pm0.85$ & $9.16\pm1.33$ \\ \hline   
\end{tabular}
\caption{Semileptonic branching ratio $B_\mu$ for different neutrino energies 
obtained from the E531 data~\cite{Bolton:1997pq} and the recent values of inclusive 
leptonic branching ratios for $D^0,D^+,D_s^+,\Lambda_{\rm c}^+$~\cite{PDG08}.      
\label{tab:BmuEnu}
}
\end{center}
\end{table}

As explained in Section~\ref{sec:freeBmu}, the uncertainty on the strange sea 
derived from the fit
can be suppressed if an additional constraint on the effective semileptonic 
branching ratio $B_\mu$ is imposed.
%
Such a constrain can come from  
a direct detection of the charmed hadrons in the emulsion experiments. The only existing 
measurement of the charmed fractions 
$f_{\rm h}$ as a function of the neutrino energy comes from a re-analysis~\cite{Bolton:1997pq} 
of the data from the E531 experiment~\cite{Ushida:1988rt,Frederiksen:1987ds}. 
Assuming $\mu$-e universality and 
the recent values~\cite{PDG08} of exclusive branching ratios for charmed hadrons 
we can determine $B_\mu$ at different neutrino energies. Our results for the E531 
data listed in Tables~\ref{tab:Bmu} and~\ref{tab:BmuEnu} correspond to 
$B_\mu(D^0)=6.53\pm0.17\%$, $B_\mu(D^+)=16.13\pm0.38\%$, $B_\mu(D_s^+)=8.06\pm0.76\%$ 
and $B_\mu(\Lambda_{\rm c}^+)=4.50\pm1.70\%$~\cite{PDG08} and take into 
account correlations among the 
measured charmed fractions~\cite{Bolton:1997pq}. Table~\ref{tab:BmuEnu} clearly shows that  
$B_\mu$ increases with energy, with more pronounced variations below 40 GeV.  
As explained in Section~\ref{sec:theory} the large contributions from quasi-elastic 
$\Lambda_{\rm c}$ and diffractive $D_s^\pm$ production at low energies explain such 
energy dependence. Potential differences between neutrinos and anti-neutrinos are 
also expected to affect mainly the region $E_\nu<40$ GeV.
This behaviour is consistent with the results of our fit to CCFR and NuTeV 
dimuon data described in Section~\ref{sec:freeBmu}.   

The CHORUS experiment also measured the production rates of charmed hadrons in 
nuclear emulsions. Thanks to a charm statistics about 20 times higher than the one 
of the E531 experiment, it was possible to directly detect some of the 
charmed-hadrons muon decays~\cite{KayisTopaksu:2005je}. The  
value of $B_\mu$ measured in Ref.~\cite{KayisTopaksu:2005je} 
is given in the first line of Table~\ref{tab:Bmu}. 
A second independent measurement of $B_\mu$ can be obtained by combining the 
inclusive charmed fractions measured in Ref.~\cite{DiCapua08} 
with the corresponding branching ratios~\cite{PDG08}, as explained above.   
The result is somewhat larger than the direct measurement as can be seen from 
the second line of Table~\ref{tab:Bmu}. 

It is worth noting that all the determinations of $B_\mu$ from emulsion 
experiments are sensitive to the value of the undetectable branching ratio 
$D^0 \to$ all neutrals (0-prongs)~\cite{KayisTopaksu:2005je,CHORUS-D0}, 
which is decreasing the overall detection efficiency.  
The recent value for the fraction of 0-prong $D^0$ decays is
$15\pm6\%$~\cite{PDG08}, 
which is intermediate between the ones assumed by the E531 and CHORUS analyses.
    
We can then proceed and average all emulsion measurements
(see Table~\ref{tab:Bmu}). Uncertainties on such averaged values of $B_\mu$ 
are smaller than the ones obtained from our fit to the CCFR and NuTeV 
dimuon data. The strange sea normalization is sensitive to 
variations of $B_\mu$, so that the inclusion of the emulsion data on $B_\mu$ to 
the fit reduces the uncertainties on the strange sea parameters.
Since the energy dependence of $B_\mu$ is more pronounced 
at small energies we use a single constraint 
$B_\mu=8.78\pm0.50\%$ for $E_\nu>30$ GeV, as an additional data point in our global fit. 
Our independent extraction of $B_\mu$ from the CCFR and NuTeV 
dimuon data, $B_\mu=9.1\pm1.0\%$, is consistent with such measurement. 
Therefore, the central value of the strange sea parameters 
obtained in this extended   
fit are comparable with those obtained if $B_\mu$ is unconstrained. 
However, the corresponding uncertainties are significantly reduced, as it can be 
seen from the second column of Table~\ref{tab:pars}. The value of the strange 
suppression 
factor becomes $\kappa(20~{\rm GeV}^2)=0.62\pm0.04$, with an uncertainty twice smaller 
as compared to the variant of the fit with $B_\mu$ unconstrained. 
With the constraint on $B_\mu$ we obtain a strange sea asymmetry $S^-=0.0013(9)$. 
This value is slightly larger than that obtained in the unconstrained fit, 
but still not significantly different from zero.

\section{Summary and Outlook}

In summary, we perform a global PDF fit using charged-lepton DIS data
on proton and deuteron,
fixed-target proton-proton and proton-deuteron Drell-Yan data,
and (anti-)neutrino induced dimuon production data from CCFR
and NuTeV experiments. 
We extract simultaneously the strange sea distributions 
and the effective semileptonic branching ratio $B_\mu$ for charmed hadrons.   
The value of $B_\mu$ obtained by our global fit is consistent with the direct 
measurements from the E531 and CHORUS emulsion experiments. 
The constraint on $B_\mu$ from emulsion data allows a reduction of the 
uncertainties on the strange sea parameters by about a factor of two.
In particular, we obtain the absolute normalization 
of the strange sea with a precision of 6\%, which is the most precise   
determination available.     
The $x$-shape of total strange sea is somewhat softer than the 
non-strange sea and the asymmetry between strange and anti-strange 
quark distributions is consistent with zero within uncertainties. 

An additional constraint on the strange sea distributions can be 
obtained from the inclusive (anti-)neutrino CC differential cross section 
$d \sigma^2_{\rm CC}/dxdy$.  
At small values of $x$ the scattering off strange sea quarks gives 
a significant contribution to the inclusive cross section.
Available cross section data come from the CHORUS~\cite{Onengut:2005kv},
NuTeV~\cite{nutev-xsec},
and NOMAD~\cite{nomad-xsec} 
experiments. 
The impact of the inclusive $\nu N$ cross sections by CHORUS  
on the strange sea distributions was recently studied in Ref.~\cite{Rojo:2008ke} 
in the context of a global PDF fit to the DIS data, resulting in a 
value of the asymmetry $S^-=-0.001\pm0.04$.  
The inclusive CHORUS data were also included in an extended 
low-$Q^2$ variant of our global PDF fit~\cite{AKP07,Alekhin:2008ua}.    

We expect a further 
improvement from the forthcoming measurements of the charmed fractions and the
inclusive charm production cross section by CHORUS~\cite{DiCapua08}. 
A global analysis 
of existing data from E531 and CHORUS emulsion experiments will allow a 
determination of $B_\mu$ at a few percent level, improving the current dominant 
source of uncertainty on strange sea distributions.  

Finally, a sample of about 15k neutrino-induced charm dimuon events is 
expected from the ongoing NOMAD analysis~\cite{NOMAD-dimu08}. These data 
were collected on an iron target with an average beam energy of 24 GeV, and 
correspond to about three times the NuTeV dimuon statistics.  
Systematic uncertainties are kept well below statistical uncertainties 
through the measurement of the ratio of dimuon to inclusive CC cross sections, 
$R_{\mu\mu}=\sigma_{\mu\mu}/\sigma_{\rm CC}$, as 
a function of different kinematic variables. Fig.~\ref{fig:rsig} shows 
a prediction for the NOMAD experiment based on our current results. 
Preliminary studies indicate that the inclusion of the NOMAD dimuon data in a global 
PDF fit would substantially reduce the uncertainties in the 
determination of the strange sea distribution. Furthermore,
%
%
an accurate measurement of $R_{\mu\mu}$ as a function of the partonic center-of-mass
energy squared $\hat{s}=Q^2\left(1/x-1\right)$ close to the charm production threshold
would allow an improved determination of the charm quark mass $m_{\rm c}$.

\section*{Acknowledgments}

This work is partially supported by the RFBR grant 06-02-16659. 
R.P. thanks USC for supporting this research.    
We are grateful to D.~Mason for providing the CCFR and NuTeV dimuon acceptance 
corrections, to A.~Mitov for stimulating 
discussions at the early stage of this work, 
to F. Di Capua for discussions and clarifications about 
the CHORUS emulsion data, to S.~Moch for the valuable comments and to S.~Mishra for 
reading the manuscript.

\end{document}